\documentclass[showpacs,preprintnumbers,
superscriptaddress,amsmath]{aipproc}

\layoutstyle{6x9}

\newcommand{\PLB}[3]{Phys.\ Lett.\ B\ {\bf #1},\ #2 (#3)}

\newcommand{\PRD}[3]{Phys.\ Rev.\ D\ {\bf #1},\ #2 (#3)}

\newcommand{\be}{\begin{equation}}
\newcommand{\ee}{\end{equation}}
\newcommand{\bea}{\begin{eqnarray}}
\newcommand{\eea}{\end{eqnarray}}
\newcommand{\ba}[1]{\begin{array}{#1}}
\newcommand{\ea}{\end{array}}

\begin{document}

\title{Light plasmon mode in the CFL phase}

\classification{12.60.Cn,12.40.Yx,14.40.Lb} 

\author{H. Malekzadeh}{
  address={Frankfurt International Graduate School for Science.} }
  
\author{Dirk H.\ Rischke}{
  address={Institut f\"ur Theoretische Physik and
Frankfurt Institute for Advanced Studies.\\
J.W.\ Goethe-Universit\"at, D-60438 Frankfurt am Main, Germany} }

\classification{}
\keywords      {}

\begin{abstract}
The self-energies and the spectral densities of 
longitudinal and transverse gluons
at zero temperature in the color-flavor-locked (CFL) phase 
are calculated. There appears a collective excitation, a light 
plasmon, at energies smaller than two times the gap parameter 
and momenta smaller than about eight times the gap. 
The minimum in the dispersion relation of this mode at 
some nonzero value of momentum corresponds to the van 
Hove singularity.
\end{abstract}

\maketitle

In cold and dense quark matter, due to asymptotic
freedom, at quark chemical potentials $\mu\ll\lambda_{QCD}$
single-gluon exchange is the dominant interaction between
quarks. Since this interaction is attractive in the color-antitriplet
channel, therefore, quark matter is a color superconductor \cite{bailinlove}.
While there are, in principle, many different color-superconducting
phases, corresponding to the different possibilities to form
quark Cooper pairs, the ground state of color-superconducting quark 
matter is the so-called color-flavor-locked (CFL) phase \cite{arw}.

At asymptotically large $\mu$, the QCD coupling constant
$g \ll 1$, thus, the gluon self-energy is dominated 
by the contributions from one quark and one gluon loop.
The quark loop is $\sim g^2 \mu^2$, while the gluon loops
are $\sim g^2 T^2$.
Since the color-superconducting
gap parameter is $\phi \sim \mu \exp(-1/g) \ll \mu$ 
\cite{son}, and since the transition temperature to the
normal conducting phase is $T_c \sim \phi$,
for temperatures where quark matter
is in the color-superconducting phase, $T$ less than $T_{c} \ll \mu$,
the gluon loop contribution can be neglected. The full description for 
the 2SC phase is given in Sec.\ II of
Ref.\ \cite{dirkigor}. The full energy-momentum dependence of 
the one-loop gluon self-energy 
has also been computed, but so far only for the 2SC phase 
\cite{2cf,dirkigor}. Here we want to do the same calculations 
for the CFL phase. The detailed computation of the 
individual components and projections can be found in the 
appendix of Ref.\cite{first}.

\begin{figure}
\includegraphics[width=9cm]{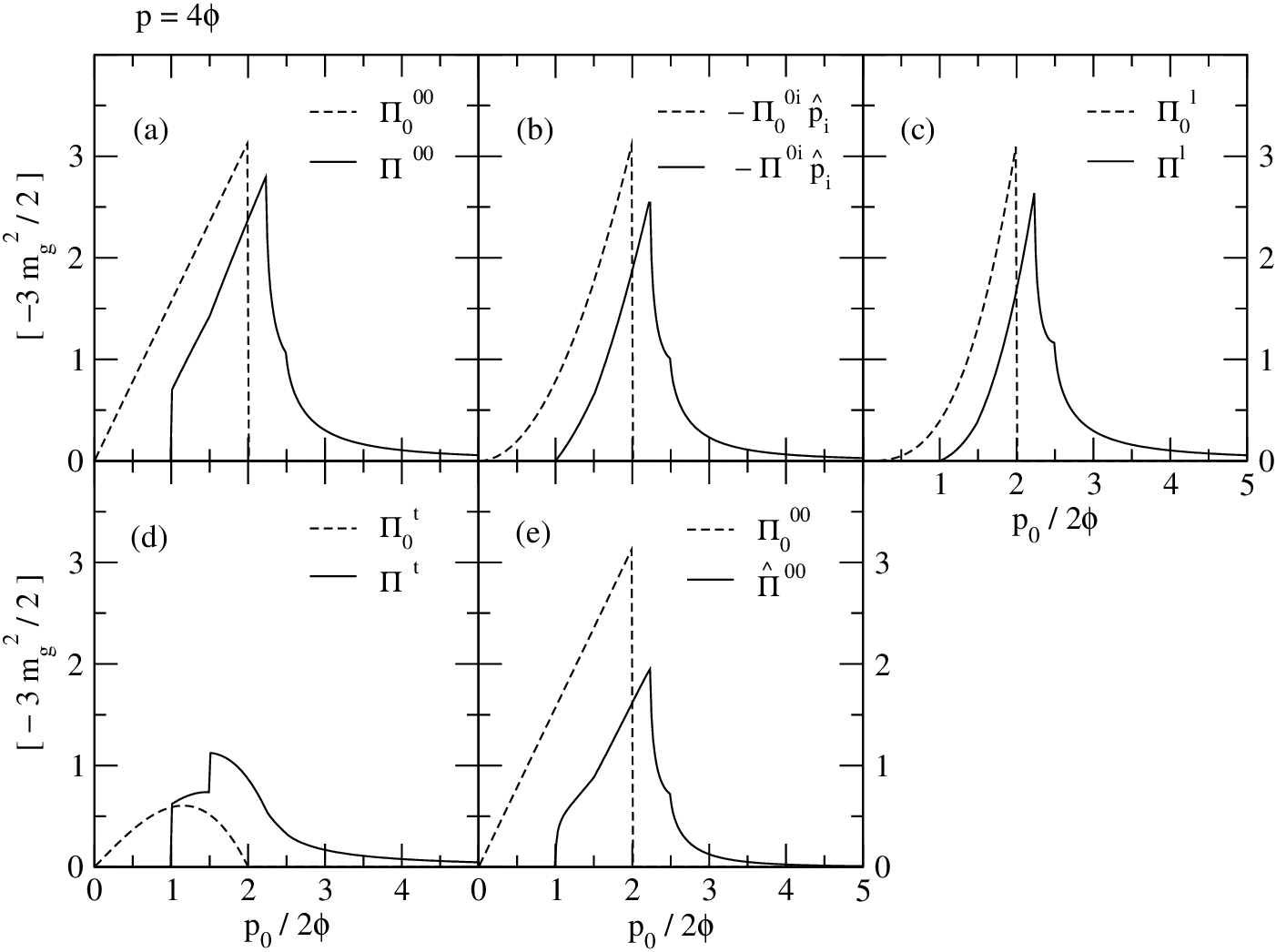}
\caption
{The imaginary parts of (a) $\Pi^{00}$, (b) $- \Pi^{0i} \hat{p}_i$,
(c) $\Pi^\ell$, (d) $\Pi^t$, and (e) $\hat{\Pi}^{00}$
as a function of energy $p_{0}$ for a gluon momentum $p=4\phi$. 
The solid lines are for the CFL phase, the dotted lines correspond
to the HDL self-energy.}
\label{im00}
\end{figure}

Fig.\ \ref{im00} shows the imaginary part of several
components of the gluon self-energy for a gluon momentum $p = 4\phi$ 
as a function of the gluon energy $p_{0}$. The corresponding results for 
the gluon self-energy in the ``hard-dense loop'' (HDL) limit, $\Pi^{\mu \nu}_0$, 
are also shown with the dotted lines.
The imaginary parts are quite similar to those of the 2SC case, 
cf.\ Fig.\ 1 of Ref.\ \cite{dirkigor}.
Nevertheless, there are subtle differences due to
appearance of two kinds of gapped quark excitations, one so-called singlet
excitation with a gap $\phi_{\bf 1}$, and eight so-called
octet excitations with a gap $\phi_{\bf 8} \equiv \phi$ \cite{arw}.
In weak coupling, the singlet gap is approximately twice as large
as the octet gap, $\phi_{\bf 1} \simeq 2\, \phi_{\bf 8} \equiv 2 \, \phi$
\cite{tomschafer1,igor}. 
Therefore, the one-loop gluon self-energy in the CFL phase has two types of
contributions, depending on whether the quarks in the loop
correspond to singlet or octet excitations, 
cf.\ Eq.\ (23b) of Ref.\ \cite{meissner3}. 
For the first type, both quarks in the loop
are octet excitations, and for the second, one is an octet and
the other a singlet excitation. 
There is no contribution from singlet-singlet excitations.

Nonvanishing octet-octet excitations require gluon
energies to be larger than $2\, \phi_{\bf 8} \equiv 2\, \phi$, 
while octet-singlet excitations
require a larger gluon energy, $p_0 \geq \phi_{\bf 1} + \phi_{\bf 8}
\equiv 3\, \phi$.
This introduces some additional structure in the imaginary parts
at $p_0 = 3 \phi$, which can be seen particularly well in Figs.\ \ref{im00}
(d) and (e). In the normal phase, the imaginary parts of the gluon self-energies 
vanish above $p_0 = p$. In color-superconducting phases, 
the imaginary parts do not vanish but fall off rapidly. This has
already been noted for the 2SC phase \cite{dirkigor}, and is
confirmed here by the results for the CFL phase.

\begin{figure}[ht]
\includegraphics[width=9cm]{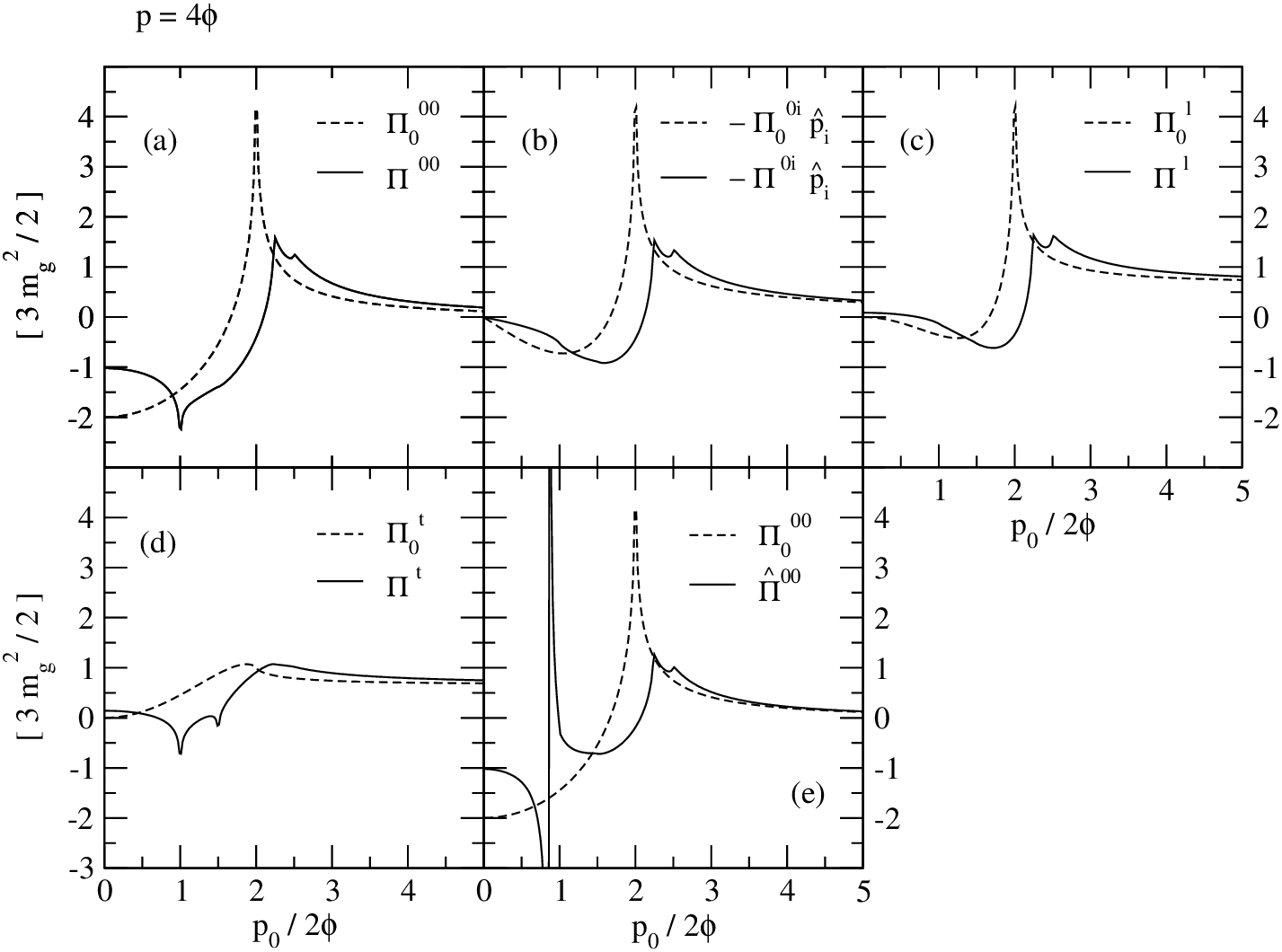}
\caption
{The real parts of (a) $\Pi^{00}$, (b) $- \Pi^{0i} \hat{p}_i$,
(c) $\Pi^\ell$, (d) $\Pi^t$, and (e) $\hat{\Pi}^{00}$
as a function of energy $p_{0}$ for a gluon momentum $p=4\phi$. 
The solid lines are for the CFL phase, the dotted lines correspond
to the HDL self-energy.}
\label{re}
\end{figure}

The real parts of the gluon self-energy are shown in Fig.\ \ref{re}.
When computing the real
part from a dispersion integral over the imaginary part, 
a change of gradient in the imaginary part
leads to a cusp-like structure in the real part. As one expects, for large energies $p_{0}\gg\phi$ 
the real parts of the self-energies 
approach the corresponding HDL limit. Deviations from the HDL limit 
occur only for gluon energies $p_{0}\sim\phi$.  

\begin{figure}[ht]
\includegraphics[width=9cm]{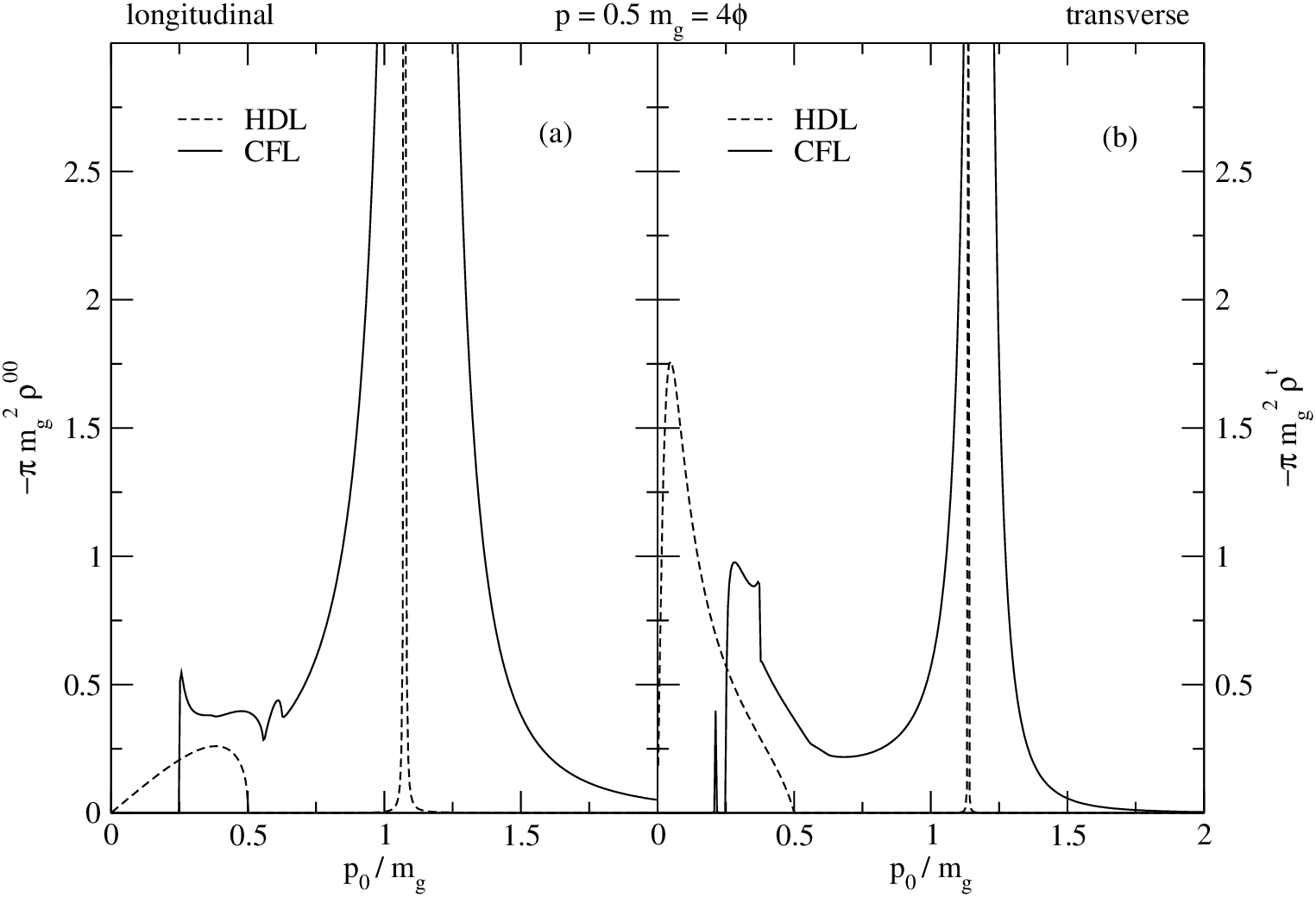}
\caption
{The spectral densities for (a) longitudinal and (b) transverse
gluons for a gluon momentum $p=m_g/2$, with $m_g = 8\, \phi$. 
The dashed lines correspond to the HDL limit.}
\label{figspecdens}
\end{figure}

\begin{figure}[ht]
\includegraphics[width=9cm]{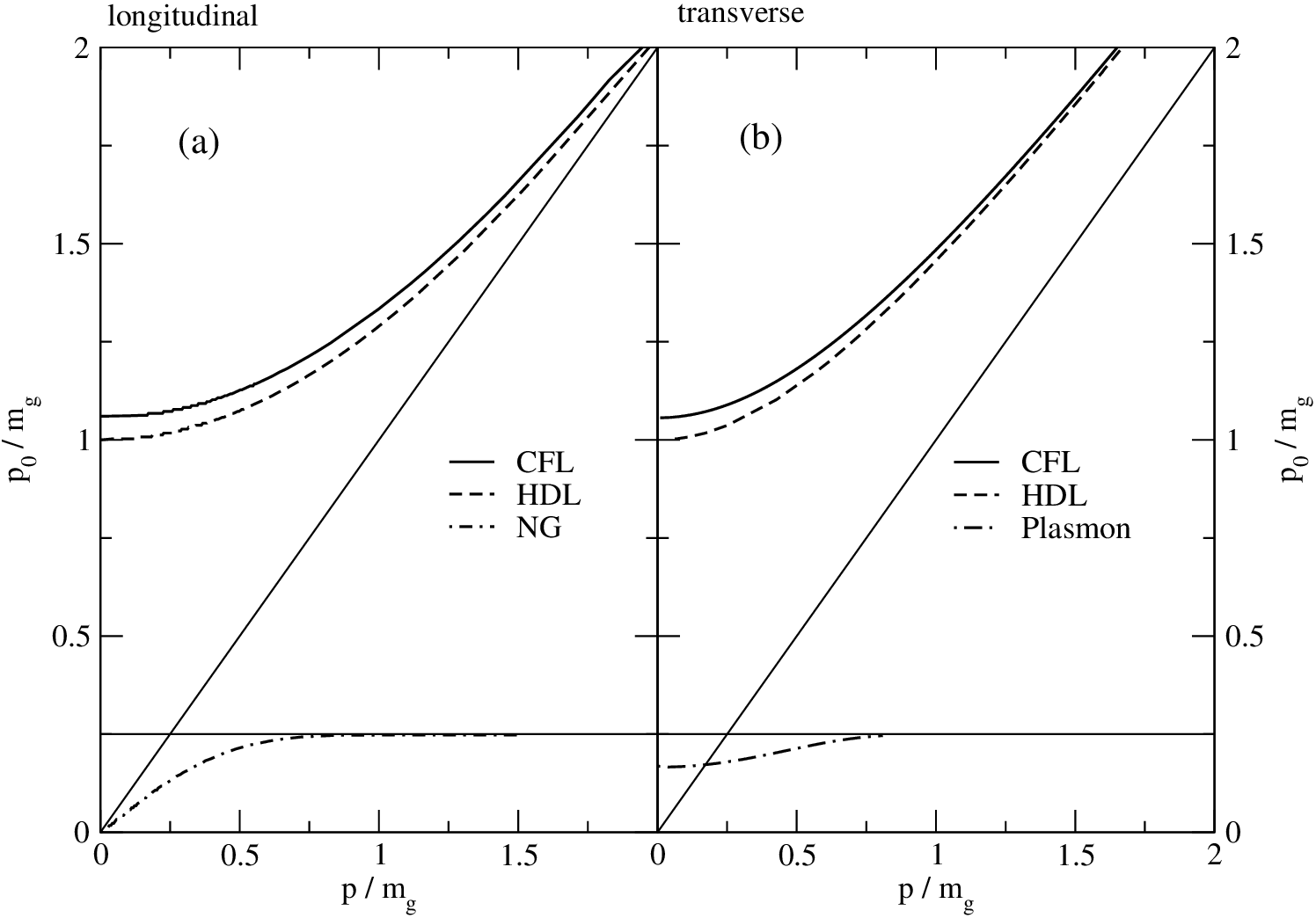}
\caption
{The dispersion relations for (a) longitudinal and (b) transverse
excitations in the CFL phase. The full lines correspond to the
regular longitudinal and transverse excitations. The
dashed lines are for the HDL limit. The dash-dotted line
in part (a) shows the dispersion relation for the Nambu-Goldstone
excitation. The light plasmon dispersion relation
is shown by the dash-dotted line in part (b). As in
Fig.\ \ref{figspecdens}, the value of the gap is chosen such that
$m_g = 8 \, \phi$.}
\label{figplasmon}
\end{figure}

The spectral densities are obtained from
the real and imaginary parts of the gluon self-energies \cite{dirkigor, first}. Note that, in Fig.\ \ref{figspecdens} at an energy $p_0 \simeq 0.21\,
m_g$, there is a delta function-like peak
in the transverse spectral density. This peak corresponds to a 
collective excitation, the so-called
``light plasmon'' predicted in Ref.\ \cite{plasmon,casalbuoni}.
We show the dispersion relation of this collective mode 
in Fig.\ \ref{figplasmon} (b). The
mass $m_{\rm coll} \simeq 1.35\, \phi$ is roughly in agreement
with the value $m_{\rm coll} \simeq 1.362\, \phi$ of Ref.\
\cite{plasmon}. As the momentum increases, the energy of the 
light plasmon excitation approaches $2\, \phi$ from below. 
For momenta larger than $\sim 8\, \phi$, the location of
this excitation branch becomes numerically indistinguishable from
the continuum in the spectral density
above $p_0 = 2\, \phi$, cf.\ Fig.\ \ref{figspecdens}.
Close inspection reveals that the dispersion relation of the
light plasmon has a minimum at a nonzero
value of $p \simeq 1.33\,\phi$, indicating a van Hove singularity.

In Fig.\ \ref{figplasmon} we also show the dispersion relations
for the ``regular'' longitudinal and transverse excitations, as well
as for the Nambu-Goldstone excitation defined by the root
of $P^\mu \, \Pi_{\mu \nu}(P)\, P^{\nu} = 0$ \cite{dirkigor,zarembo}.
For our choice of gauge, the
gluon propagator is 4-transverse and this mode does not
mix with the longitudinal
component of the gauge field \cite{dirkigor}. 
Therefore, the Nambu-Goldstone
mode does not appear as a peak in the
longitudinal spectral density, cf.\ Fig.\ \ref{figspecdens}.
We finally note that other collective excitations have been
investigated in Ref.\ \cite{iida}.

In conclusion, we have computed the gluon self-energy 
in the CFL phase as a function of energy and momentum. While the
imaginary parts of the gluon self-energy could be expressed
analytically in terms of elliptic functions (see appendix of \cite{first}), the real parts
had to be computed numerically with the help of dispersion
integrals. From 
the real and imaginary parts we constructed the spectral
densities. We confirmed the existence of a low-energy collective
excitation, the so-called ``light plasmon'' predicted in Ref.\
\cite{plasmon}.

\section{Acknowledgments}
H.\ M.\ thanks the Frankfurt International Graduate School of Science
for support.

\end{document}